\documentclass[reprint,aps,pra,twocolumn,superscriptaddress]{revtex4-1}
\usepackage{braket}
\usepackage{amssymb}
\usepackage{amsmath}
\usepackage{epsfig}
\usepackage{color}
\usepackage{graphics, graphicx}
\usepackage{bbold}
\usepackage{psfrag}
\usepackage{mathcomp}
\usepackage{subfigure}
\usepackage{float}
\usepackage{verbatim}
\usepackage[colorlinks, citecolor=blue, linkcolor=blue]{hyperref}
\usepackage[normalem]{ulem}
\usepackage{bm}

\begin{document}

\title{Self-bound vortex lattice in a rapidly rotating quantum droplet}
\author{Qi Gu}
\affiliation{Institute for Advanced Study, Tsinghua University, Beijing 100084, China}
\author{Xiaoling Cui}
\email{xlcui@iphy.ac.cn}
\affiliation{Beijing National Laboratory for Condensed Matter Physics, Institute of Physics, Chinese Academy of Sciences, Beijing 100190, China}

\date{\today}

\begin{abstract}
A rapidly rotating Bose gas in the quantum Hall limit is usually associated with a melted vortex lattice. 
In this work, we report a self-bound and visible triangular vortex lattice without melting for a two-dimensional Bose-Bose droplet rotating in the quantum Hall limit, i.e., with rotation frequency $\Omega$ approaching the trapping frequency $\omega$.
Increasing $\Omega$ with respect to interaction strength $U$, we find a smooth crossover of the vortex lattice droplet from a {\it needling} regime, as featured by small vortex cores and an equilibrium flat-top surface, to the lowest-Landau-level regime with Gaussian-extended cores spreading over the whole surface.   
The surface density of such a rotating droplet is higher than that of a static one, and their ratio is found to be a universal function of $\Omega/U$. We have demonstrated these results by both numerical and variational methods. The results pave the way for future experimental exploration of rapidly rotating ultracold droplets into the quantum Hall limit.  
\end{abstract}
\maketitle

\section{Introduction}
Ultracold Bose gases in rotating harmonic traps have attracted much attention in recent years  \cite{review1,review2}. An important motivation is the emergence of quantum Hall physics when the rotating frequency ($\Omega$) approaches the trapping frequency ($\omega$)  \cite{JasonHo, Cooper}. This situation is analogous to charged particles in a magnetic field, and a triangular vortex lattice emerges therein similar to  those  in type-II superconductors \cite{Abrikosov1957} and rotating superfluid heliums \cite{Tkachenko1966}. In such a rotating dilute gas, the competition between inertial force, trapping potential and mean-field interaction results in a rich profile of vortex lattice, such as distinct core and global envelopes between mean-field quantum Hall and Thomas-Fermi regimes \cite{Fischer2003,Baym2004} and structural transitions to other lattice configurations in two-species bosons \cite{Mueller_Ho2002,Ueda2003,Cornell2}, dipolar systems \cite{Cooper2,HuiZhai1,Kumar2017}, Bose-Fermi double superfluids \cite{HuiZhai2}, etc.  
Despite these interesting phenomena, in practice it is quite  challenging  to explore the vortex lattice in the rapidly rotating quantum Hall regime. A key obstacle is the exceedingly expanded cloud due to the cancellation between centrifugal and confinement forces ($\Omega=\omega$). The resulted low density of the cloud leads  to the melting of vortex lattice, which shows an invisible pattern \cite{QH_expt1, QH_expt2} and finally gives way to other correlated states \cite{Cooper, Sinova2002:QuantumMeltingAbsence,Regnault2003:QuantumHallFractions}. Alternatively, the boson density could be enhanced in the Landau-gauged quantum Hall regime by 
squeezing the system with anisotropic traps \cite{MIT_expt1, MIT_expt2}, which  displays a crystalline vortex street \cite{Gora} instead of a triangular lattice as in the symmetric case.

Here, we show that the above obstacle can be overcome in the quantum droplet of ultracold bosons. Such an ultracold droplet, as balanced by the mean-field attraction and Lee-Huang-Yang(LHY) repulsion from quantum fluctuations \cite{Petrov2015}, has been successfully realized in both dipolar gases \cite{Ferrier-Barbut2016,Schmitt2016,Chomaz2016, Modugno,Pfau_4,Ferlaino_2} and alkali Bose-Bose mixtures \cite{Cabrera2018,Cheiney2018,Semeghini2018,DErrico2019,  DJWang}. Importantly, these droplets are self-bound and can be stabilized without any trap. They are thus expected to be immune from expansion as in previous cases of rotating Bose gases \cite{QH_expt1, QH_expt2}.
Although there have been several studies on a few vortices in dipolar droplets \cite{Cidrim2018:VorticesSelfboundDipolar,Roccuzzo2020:RotatingSupersolidDipolar,Ancilotto2021:VortexPropertiesExtended,Klaus2022:ObservationVorticesVortexa,Li2023:AnisotropicVortexQuantum,Bland2023:VorticesDipolarBoseEinstein} and binary boson droplets \cite{Li2018:TwodimensionalVortexQuantum,Zhang2019:SemidiscreteQuantumDroplets,Tengstrand2019:RotatingBinaryBoseEinstein,Caldara2022:VorticesQuantumDroplets,Kuan2023:DynamicsRapidlyRotating,Nikolaou2023:NovelSuperfluidStates}, the property of a rapidly rotating droplet in the quantum Hall limit is still unknown at the moment. In particular, how the self-bound nature and beyond-mean-field effect will influence the vortex lattice is an important question to address.

\begin{figure}[t]
    \centering
    \includegraphics[width=8.5cm]{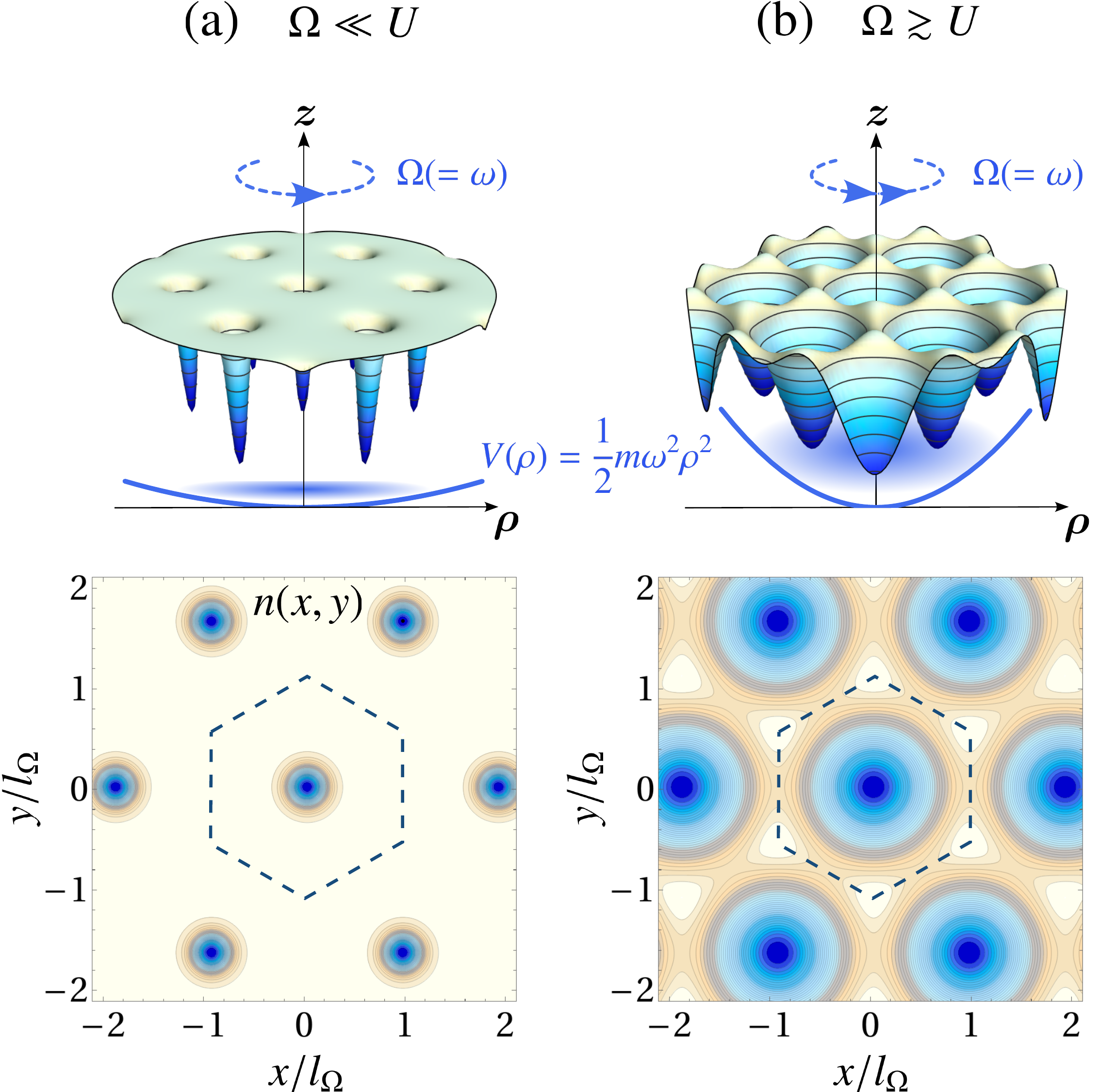}
    \caption{Two limiting cases of vortex lattice for a 2D quantum droplet at different $\Omega/U$. Here $\Omega$ is the rotation frequency ($=$ trapping frequency $\omega$) and $U$ measures the interaction strength. For small $\Omega/U=0.008$ (a), each unit cell exhibits a rather small vortex core surrounded by a large flat-top area, resembling a needled surface. Increasing $\Omega/U=3.085$ (b), the system is frozen at the lowest Landau level, where the vortex core displays an extended Gaussian profile and the surface area becomes isolated and small. The dashed hexagon in each $n(x,y)$ plot marks a unit cell with characteristic length $l_{\Omega}=1/\sqrt{m\Omega}$. 
}\label{fig_schematic}
\end{figure}

In this work, we study the vortex lattice in a harmonically trapped two-dimensional (2D) Bose-Bose droplet with extreme rotation $\Omega=\omega$. Our calculations confirm that 
the self-bound nature of the quantum droplet indeed protects the vortex lattice from melting, thus enabling a clear visualization of the triangular lattice in this limit. As shown in Fig.~\ref{fig_schematic}, the exact vortex structure in a unit cell closely depends on the relative strength between $\Omega$ (or $\omega$) and interaction strength $U$. Namely, at $\Omega\ll U$ [Fig.~\ref{fig_schematic}(a)], each vortex holds a rather small core with a large flat-top area around, such that the entire system resembles a needled surface. Increasing $\Omega/U$ will finally drive the system to the lowest-Landau-level (LLL), where the vortex core displays an extended Gaussian profile across the whole cell and the surface area becomes isolated and small [Fig.~\ref{fig_schematic}(b)].
We emphasize that the flat-top surface here 
inherits from the static droplet in vacuum and  reflects the crucial role played by LHY corrections. 
Interestingly, the surface density here is always higher than that of a static droplet, and their ratio universally relies on a single parameter $\Omega/U$. These results have been demonstrated both numerically and from a variational approach. They will hopefully serve as a guideline for  future
experimental exploration of rapidly rotating ultracold droplets in the quantum Hall limit.

\section{Model}
We start from the energy functional of two-species bosons in a 2D rotating trap ($\hbar=1$):
\begin{eqnarray}
\cal{E}(\bm{\rho})&=&-\sum_{i=1,2}  \phi_i^{*}(\bm{\rho})\left( -\frac{\nabla^2}{2m} + \frac{m\omega^2 \bm{\rho}^2}{2} - \Omega L_z\right) \phi_i(\bm{\rho}) \nonumber\\
&&+ \sum_{ij}\frac{g_{ij}}{2}n_i(\bm{\rho})n_j(\bm{\rho})+ {\cal E}_{\rm LHY}[n_1(\bm{\rho}),n_2(\bm{\rho})];\label{E}
\end{eqnarray}
here $\bm{\bm{\rho}}=(x,y)$ is the 2D coordinate; $\phi_i$ is the wavefunction of the $i$-th species and $n_i=|\phi_i|^2$ is its density; $g_{ij}=4\pi a_{ij}/(ml_z)$ is the mean-field coupling between $i$- and $j$ species ($a_{ij}$ is the scattering length and $l_z$ is the confinement length along $z$ to generate 2D geometry); and $L_z=-i (\bm{\rho}\times \nabla)_z $ is the angular momentum's $z-$component. In this work we shall consider bosons near the mean-field collapse, i.e., with $\delta g\equiv g_{12}+ \sqrt{g_{11}g_{22}}\sim 0$. In this case, the two-species densities are locked as $n_1/n_2=\sqrt{g_{22}/g_{11}}$ in order to minimize the mean-field energy \cite{Petrov2015}. As a result, one can employ the single-mode wave function
\begin{equation}
\Psi(\bm{\rho})=\left( \frac{\sqrt{g_{11}}+\sqrt{g_{22}}}{\sqrt{g_{22}}}\right)^{1/2} \phi_1 = \left( \frac{\sqrt{g_{11}}+\sqrt{g_{22}}}{\sqrt{g_{11}}}\right)^{1/2} \phi_2,
\end{equation}
and simplify (\ref{E}) as
\begin{eqnarray}
\cal{E}(\bm{\rho})&=&\Psi^{*}\left( -\frac{\nabla^2}{2m} + \frac{m\omega^2 \bm{\rho}^2}{2} - \Omega L_z\right)  \Psi + \frac{g}{2}n^2+ {\cal E}_{\rm LHY}(n). \nonumber\\ \label{E2}
\end{eqnarray}
Here, $n=|\Psi|^2$ is the total density, the reduced mean-field coupling is $g=2\delta g \sqrt{g_{11}g_{22}}/(\sqrt{g_{11}}+\sqrt{g_{22}})^2$, and the LHY energy reads ${\cal E}_{\rm LHY}=an^2 \ln(bn)$, with $a=2\pi a_{11}a_{22}/(ml_z^2)$ and $b=4\pi l_z \sqrt{ea_{11}a_{22}}$ \citep{Petrov_2, Zin}, with $e$ the Napier's constant. The ground state  can then be  obtained by imaginary-time evolution of the Gross-Pitaevskii(GP) equation $i\partial_t\Psi=\partial {\cal{E}}/\partial \Psi ^*$ based on (\ref{E2}). 

For a concrete example we shall take the two  hyperfine states of ${}^{39}$K atoms, $\ket{1} = \ket{F=1,m=-1}$ and $\ket{2}=\ket{F=1,m=0}$, as have been well studied in droplet experiments  \cite{Semeghini2018,Cabrera2018,Cheiney2018}. Then we have $a_{11}=35a_B$ and $a_{12}=-53a_B$ ($a_B$ is the Bohr radius) and  $a_{22}$ is highly tunable through Feshbach resonance. Moreover, we assume a tunable $\Omega(=\omega)$ as well, while we fix $l_z$ as $0.05\mu$m. 
In our numerics, a large boson number $N>10^4$ is considered, such that the droplet stays in thermodynamic limit with little finite-size effect. The boson filling factor $\nu=N/N_v$, with $N_v$ the vortex number, is $> 150$, well above the critical value for the instability of vortex lattice \cite{Cooper,Sinova2002:QuantumMeltingAbsence,Baym2004_modes}. Such large $\nu$ in turn validates the matter wave treatment in the GP equation.  To facilitate later discussions, we introduce  a length scale $l_{\Omega}=1/\sqrt{m\Omega}$ to characterize the size of a unit cell, and an energy scale 
\begin{equation}
U=an_0
\end{equation}
to characterize typical interaction strength, where  $n_0=b^{-1}e^{-1-g/(2a)}$ is the equilibrium density of a static droplet in vacuum.

We have numerically obtained the ground state of rapidly rotating ${}^{39}$K droplets at $\Omega=\omega$ (see details presented in the Appendix).  For all values of $\Omega/U$, we find a stable and  regularly distributed triangular vortex lattice without melting, in contrast to those in repulsive Bose gases in the same quantum Hall limit \cite{QH_expt1, QH_expt2}. The vortex lattice is thus intrinsically self-bound, which will greatly facilitate its detection in experiments. Its internal structure, as summarized in Fig.\ref{fig_schematic},  falls into two limiting regimes depending on the ratio $\Omega/U$:

{\bf I. Needling regime:} when $\Omega\ll U$, the vortex core is very small as compared to the size of unit cell, and the rest large area is filled with bosons with flat-top (equilibrium) density. This structure, as shown in Fig.\ref{fig_schematic}(a), resembles a needled surface and thus it is called the needling regime.

{\bf II. LLL regime:} when $\Omega\gtrsim U$, the vortex spreads to the whole cell and the surface becomes isolated and individually small, see Fig.\ref{fig_schematic}(b). As shown later, the vortex profile in this case saturates at an extended Gaussian (Fig.\ref{fig_LLL}), a characteristic feature of the LLL regime.

\section{Variational approach}
To physically understand these different structures, we write down the wave-function of the rotating droplet:
\begin{equation} 
    \Psi(\bm{\rho})=e^{i \Phi(\bm{\rho})}f(\bm{\rho})\sqrt{\bar{n}}, \label{wf}
\end{equation}
where $\Phi(\bm{\rho})$ is the phase, $f(\bm{\rho})$ is a real function rapidly varying on the scale of the vortex core, and $\bar{n}$ is the equilibrium density far from the core. Since $f(\bm{\rho})$ has the same discrete translational symmetry as the vortex lattice, we can divide the lattice into Wigner-Seitz cells and treat $f$ in each cell separately. Then the vortex number is $N_v=\sum_j$ ($j$ is the cell index), and the total boson number is $N= N_v \nu$, with $\nu=\bar{n} \int_j d\bm{\rho} f^2$  the filling factor (here $\int_j$ is to integrate within a unit cell).

Now we aim to  express the total energy $E=\int d{\bm{\rho}} {\cal{E}}(\bm{\rho}) = E_0+E_{\rm int}$  in terms of $f$ and $\bar{n}$ in (\ref{wf}), where $E_0$ is the non-interacting energy and $E_{\rm int}$ is the interaction energy. Since each vortex is singly quantized, we introduce the velocity field as $\bm{v}(\bm{\rho})\equiv \nabla \Phi/m$, which is contributed from an overall rigid rotation and a periodic local velocity, i.e., $\bm{v}(\bm{\rho})=\Omega \bm{e}_z\times \bm{\rho} + \bm{v}_l(\bm{\rho})$ \cite{Tkachenko1966}. After straightforward algebra, we simplify  $E_0$ as
\begin{eqnarray}
E_0&=&N\Omega \alpha; \label{E0} \\
{\rm with} \ \    \alpha &=& \frac{ \int_j d\bm{\rho} \left( \frac{1}{2m}\big(\frac{\partial
     f}{\partial \rho}\big)^2 + \frac{m}{2} f^2 \bm{v}_l^2\right)}{ \Omega \int_j d\bm{\rho}f^2}. \nonumber
\end{eqnarray}
To exactly integrate  $\bm{v}_l^2$ in $\alpha$, we adopt a complex function $v_l(z) = i (\zeta^*(z)/m - \Omega z)$ in terms of $z=x+iy$, where $\zeta(z)$ is the Weierstrass zeta function
$
    \zeta(z) = 1/{z} + \sum_{j\ne0}\big[{1/(z-z_j)} + 1/{z_j} + {z}/{z_j^2}\big],
$
and $\{z_j\}$ are complex coordinates of lattice sites  \cite{Akhiezer1990:ElementsTheoryElliptic}. 
Clearly $\alpha$ depends on the symmetry of the vortex lattice, and its minimization  leads to the emergence of triangular lattice as shown later.

The interaction energy, as composed by mean-field and LHY parts, is given by
\begin{eqnarray}
&&\ \ \ \ \ \   E_{\rm int}=N\beta\Big(g\bar{n}/2 + a \bar{n}\ln(b\bar{n})\Big) + N\gamma a \bar{n}; \label{Eint} \\
&&{\rm with}  \ \ \ \ \beta = \dfrac{\int_j d\bm{\rho}f^4}{{\int_j d\bm{\rho}\, f^2}}, \ \ \ \  \gamma = \dfrac{\int_j d\bm{\rho}f^4\ln f^2}{{\int_j d\bm{\rho}\, f^2}}. \nonumber
\end{eqnarray}
The total energy per particle is then $\epsilon=E/N$:
\begin{equation}
\epsilon=\alpha\Omega+ \beta\big(g\bar{n}/2 + a \bar{n}\ln(b\bar{n})\Big) + \gamma a \bar{n}. \label{total_E}
\end{equation}
To this end, we have obtained $\epsilon$ as a function of $\bar{n}$ and three dimensionless quantities $\alpha,\ \beta,\ \gamma$ that are uniquely determined by $f$-function. Here we remark that the inclusion of LHY energy in $\epsilon$ is essential to support a self-bound solution at zero pressure, which is equivalent to requiring $\partial \epsilon/\partial \bar{n}=0$ and gives 
\begin{eqnarray}
    \bar{n}&=&e^{-\gamma/\beta}n_0, \label{nbar} \\
    \epsilon &=& \alpha\Omega - \beta e^{-\gamma/\beta}U.  \label{epsilon}
\end{eqnarray}
Here $n_0$ and $-U$ are, respectively, the equilibrium density and energy per particle for a static droplet in vacuum. 

To further parametrize $\alpha,\ \beta$, and $\gamma$, we now introduce a variational ansatz for $f(\rho)$ ($\rho=|\bm{\rho}|$) within a unit cell. Considering its definition in (\ref{wf}) with boundary condition $f(\rho\rightarrow 0)\propto \rho$, we write down the ansatz as 
\begin{equation}\label{trial_f}
    f({\rho})=\begin{cases}
        (\rho/l_c)\ e^{1-\rho/l_c}, & \rho\le l_c, \\
        1, & \rho> l_c.
    \end{cases} 
\end{equation}
Here, $l_c$ characterizes the core radius, and one can check that both $f$ and $f'$ are continuous at $\rho=l_c$. Define a dimensionless quantity $\eta=l_c^2/l_{\Omega}^2$, which evaluates the core area with respect to the area of a unit cell, we can then parametrize $\alpha,\beta,\gamma$ solely by $\eta$. The ground state is then given by $\partial\epsilon /\partial \eta=0$, which leads to
\begin{equation}
    \dfrac{\Omega}{U}  \dfrac{\partial \alpha}{\partial \eta} =  \dfrac{\partial(\beta e^{-\gamma/\beta})}{\partial \eta}. \label{eta_eq}
\end{equation}
Importantly, Eq.~\eqref{eta_eq} implies that the solution of $\eta$, and thus other dimensionless quantities including  $\{\alpha,\beta,\gamma\}$ and the ratio $\bar{n}/n_0$, all rely on a single parameter $\Omega/U$. This parameter, according to Eq.(\ref{epsilon}), measures the relative strength between non-interaction and interaction energy scales.

\begin{figure}[!ht]
    \centering
    \includegraphics[width=8cm]{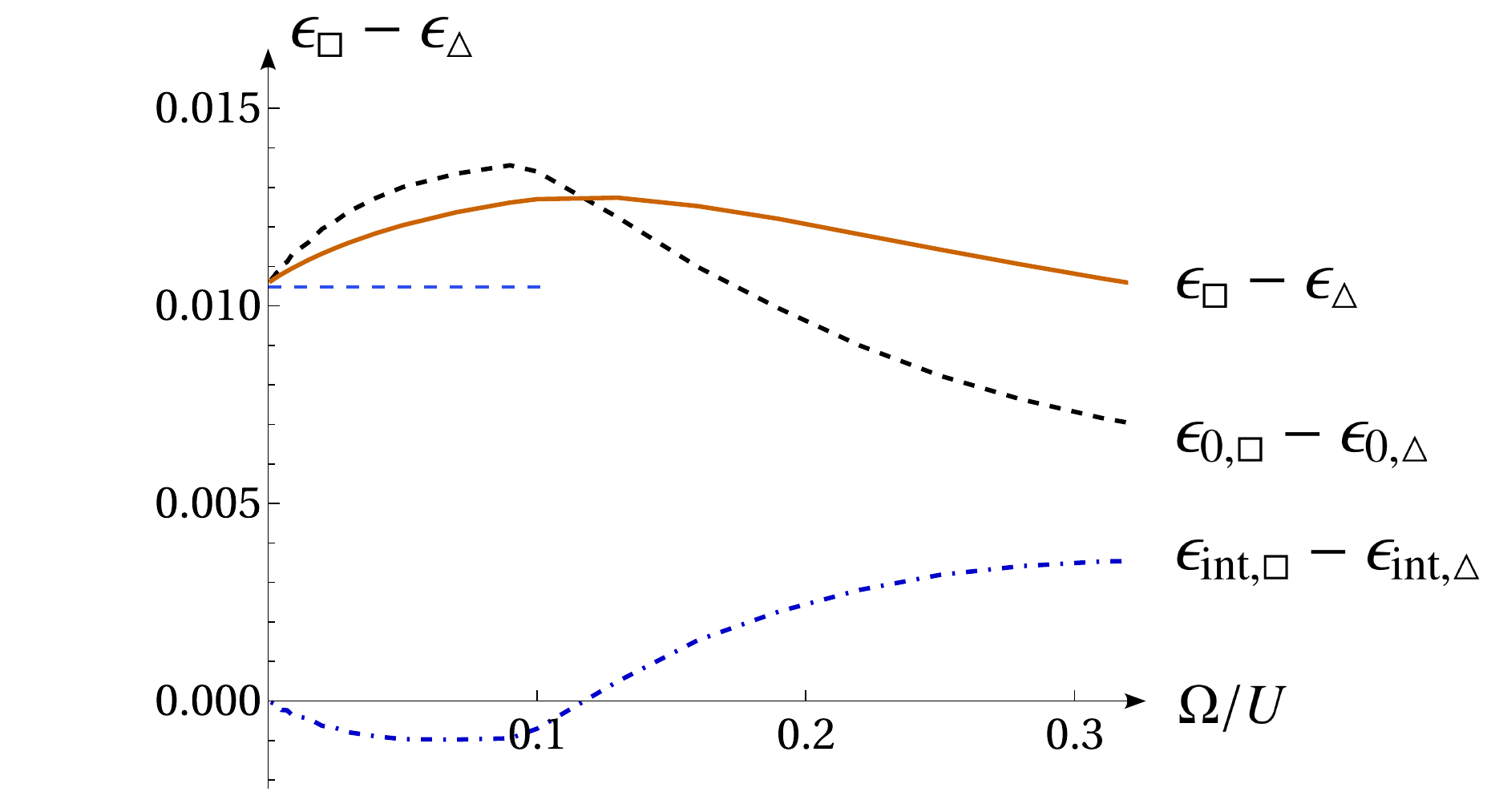}
    \caption{Difference of energy per particle between the square and triangular vortex lattices $\epsilon_{\square}-\epsilon_{\bigtriangleup}$. The contributions from the non-interaction and interaction parts are  shown as $\epsilon_{0,\square}-\epsilon_{0,\bigtriangleup}$ and $\epsilon_{{\rm int},\square}-\epsilon_{{\rm int},\bigtriangleup}$, respectively. The horizontal dashed line marks the value obtained in incompressible limit  \cite{Tkachenko1966}. The energy unit here is the $\Omega$. }
    \label{fig_triangle}
\end{figure}

\section{Results and discussion}
The above variational approach can well explain the structure of vortex lattices as shown in  Fig.~\ref{fig_schematic}. First, it  gives the triangular lattice as the ground state configuration for all values of $\Omega/U$. To see this, in Fig.~\ref{fig_triangle} we have shown the energy difference between the square and triangular lattices based on Eqs.~\eqref{epsilon} and \eqref{eta_eq}. The energy per particle $\epsilon$ is divided into non-interacting part $\epsilon_0$ and interacting part $\epsilon_{\rm int}$. We use subscripts $\square$ and $\bigtriangleup$ to distinguish square and triangle configurations. One can see that for all $\Omega/U$, the triangular lattice is always more energetically favorable, and the energy gain is mostly contributed from the noninteracting part, or equivalently, the difference in $\alpha$. In the incompressible limit with $\Omega/U,\eta\rightarrow 0$, our calculation {(the difference in $\alpha$)} reproduces Tkachenko's result with $\frac{m^2}{2\pi}\left(\int_{\square}v_{l}^2-\int_{\bigtriangleup }v_{l}^2 \right)\approx 0.0105 $ \cite{Tkachenko1966}.

\begin{figure}[!ht]
    \centering
    \includegraphics[width=\linewidth]{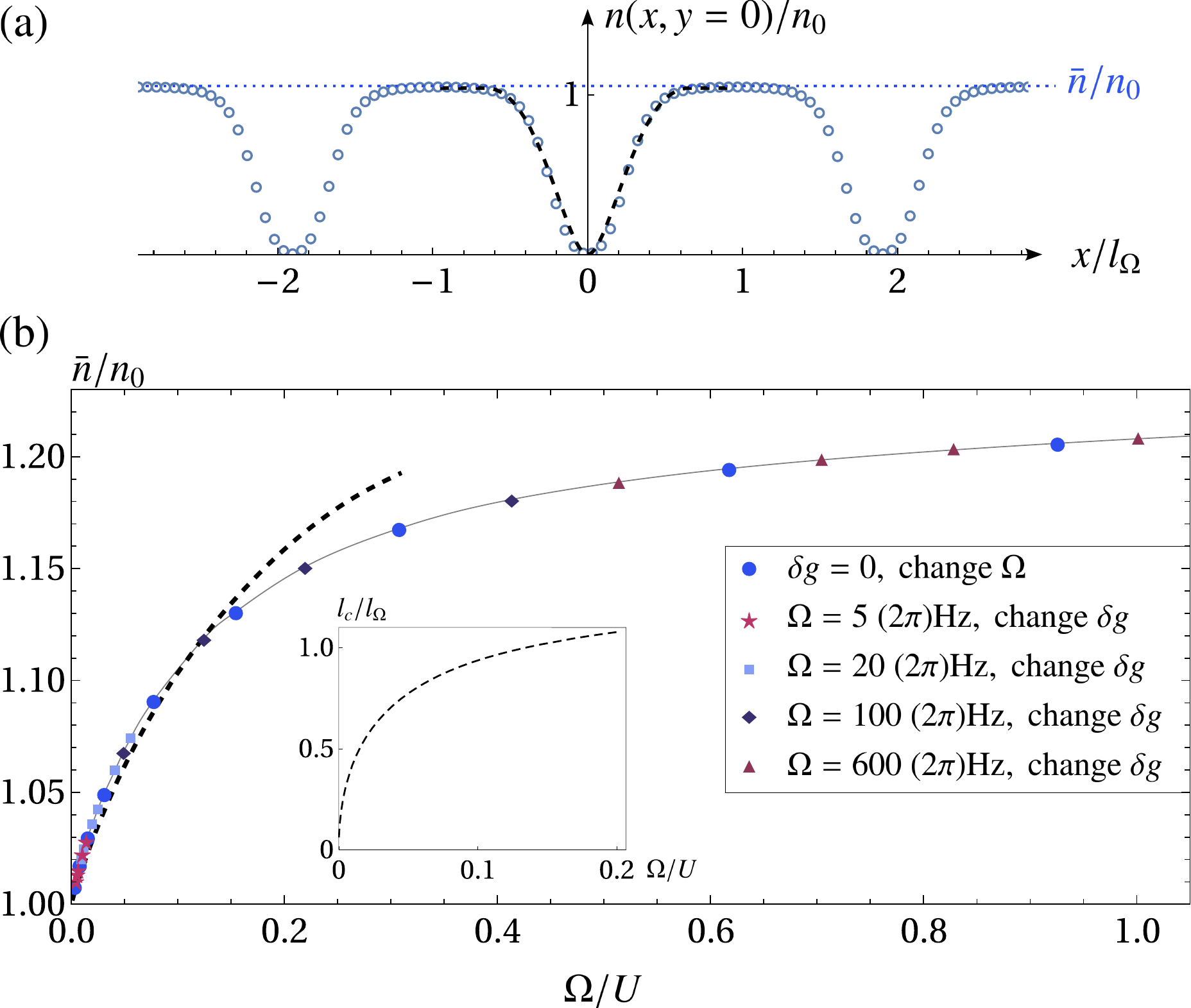}
    \caption{Vortex lattice in the needling regime with $\Omega\ll U$. (a) Density profile $n(x,y=0)$  across three vortex cores at $\Omega/U=0.03$. Dashed line shows the variational function fit according to Eq.~\eqref{trial_f} with optimized core radius $l_c$. 
    (b) $\bar{n}/n_0$ as a function of $\Omega/U$, with $\bar{n}$ (or $n_0$) the equilibrium surface density of a rotating droplet (or a static droplet in vacuum). Here, $\Omega/U$ is changed  by varying different parameters ($\Omega$ or $\delta g$) and all data collapse into a single  curve (gray solid line),  signifying a {\it universal} dependence of $\bar{n}/n_0$ on $\Omega/U$. Black dashed line shows the results of the variational approach based on Eq.~\eqref{trial_f}, and the according $l_c/l_{\Omega}$ is shown in the inset plot.}
    \label{fig_needling}
\end{figure}

Secondly, Eqs.~\eqref{total_E}-\eqref{eta_eq} well explain the change of vortex core structure as tuning $\Omega/U$.
Let us start from the needling regime at $\Omega\ll U$. In this case, since the interaction part dominates in $\epsilon$, the system is largely unperturbed with an equilibrium density $\bar{n}$, while just a small region in each cell is significantly deformed by the vortex. In Fig.~\ref{fig_needling}(a), we plot out a typical density profile for this case and show that it can be well fit using the variational ansatz in \eqref{trial_f}.
In Fig.~\ref{fig_needling}(b), we further extract $\bar{n}/n_0$ as a function of $\Omega/U$. The data show that $\bar{n}$ for a rotating droplet is {\it always} higher than $n_0$ for a static one, and their ratio $\bar{n}/n_0$ gradually increases with $\Omega/U$. This can be attributed to the factor $e^{-\gamma/\beta}>1$ in (\ref{nbar}), given $\gamma<0$ according to its definition. The  variational approach is found to provide a quantitatively good prediction to $\bar{n}/n_0$ for small $\Omega/U\lesssim 0.15$, and it also predicts an increasing core size in each unit cell, see $l_c/l_{\Omega}\sim \Omega/U$ in the inset of Fig.~\ref{fig_needling}(b). Remarkably, by choosing different variables ($\Omega$ or $\delta g$) in changing $\Omega/U$, we find that all data of $\bar{n}/n_0$ collapse into a single curve in Fig.\ref{fig_needling}(b). This demonstrates a {\it universal} dependence of $\bar{n}/n_0$ on $\Omega/U$, as suggested previously in the variational approach [see Eq.~\eqref{eta_eq}]. 

\begin{figure}[!ht]
    \centering
    \includegraphics[width=\linewidth]{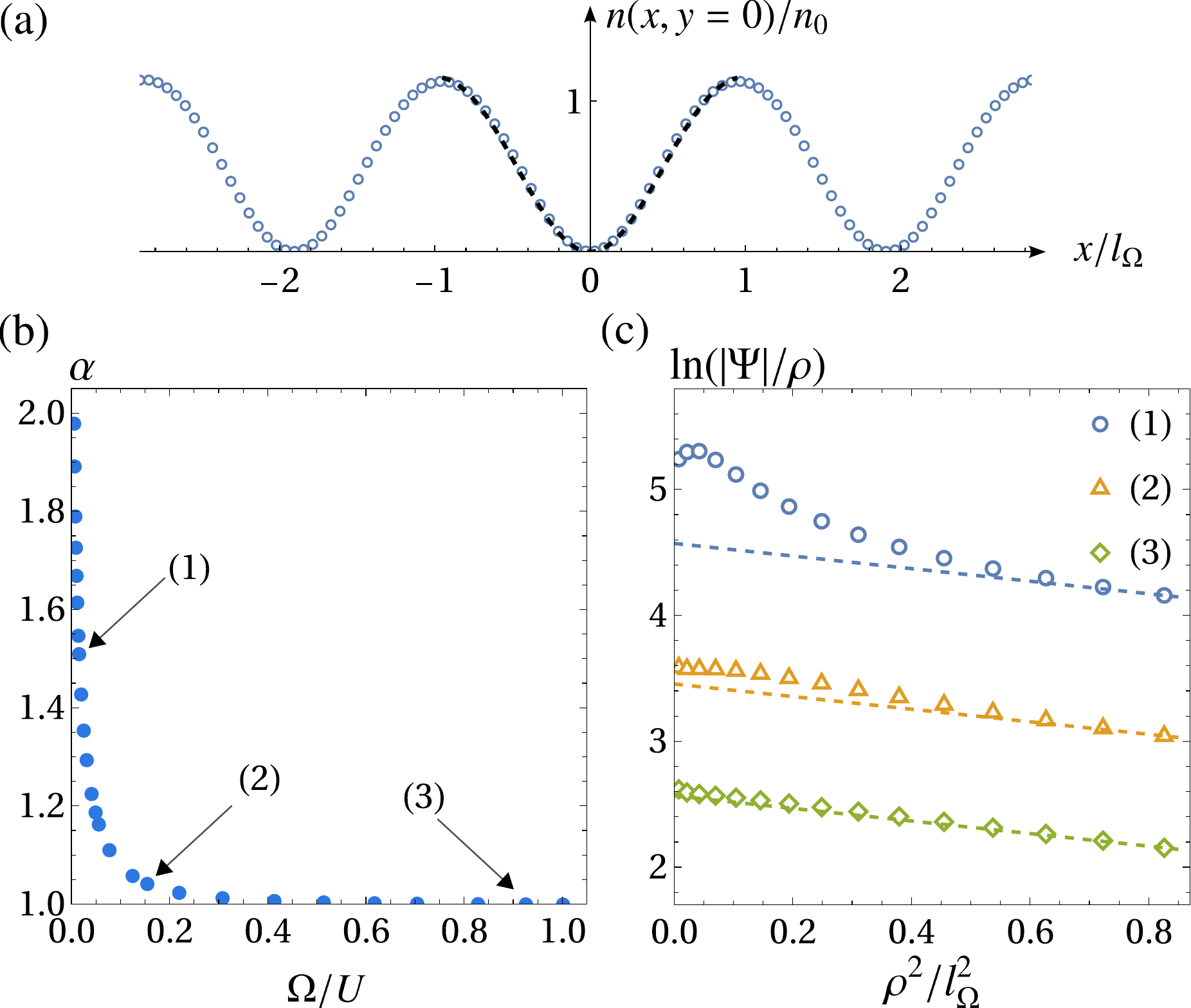}
    \caption{Crossover of vortex lattice from the needling to the LLL regime. (a) Same as Fig.\ref{fig_needling}(a) except for  $\Omega/U=1$. Dashed line shows the Gaussian function (\ref{Gaussian}). (b)  $\alpha=E_0/(N\Omega)$ as a function of $\Omega/U$.  (c) $y=\ln(|\Psi|/\rho)$ as a function of $x=\rho^2/l_{\Omega}^2$ for different $\Omega/U=0.02,\ 0.15$, and $0.92$, as marked by (1),(2), and (3) in (b). For case (3), all points collapse into a straight line $y=-x/2+c$ (see dashed fit), which justifies the LLL wave function in \eqref{Gaussian}.  }
    \label{fig_LLL}
\end{figure}

For $\Omega/U\gtrsim 0.15$,  the system no longer stays in the needling regime, as indicated by a considerably large core in a unit cell ($l_c/l_{\Omega}\sim 1$) and the breakdown of variational ansatz in predicting $\bar{n}/n_0$, see Fig.\ref{fig_needling}(b). In this case, the system gradually evolves to the LLL regime, and the energies in both Eqs.~\eqref{E2} and \eqref{total_E} are dominated by the non-interacting parts, i.e., with noninteracting $\epsilon_0$ approaching the LLL energy $\Omega$ (thus $\alpha\rightarrow 1$). The associated $f(\rho)$ then develops an extended Gaussian distribution 
\begin{equation}
    f({\rho})=\begin{cases}
        (\rho/l_\Omega)\ e^{1/2-\rho^2/(2l_\Omega^2)}, & \rho\le l_\Omega, \\
        1, & \rho> l_\Omega.
    \end{cases} .
    \label{Gaussian}
\end{equation}
Figure.~\ref{fig_LLL}(a) shows that the above function indeed determines the actual density within a unit cell at a large $\Omega/U=1$.

The gradual crossover from the needling to LLL regimes is shown in Figs.~\ref{fig_LLL}(b) and \ref{fig_LLL}(c), where we have extracted $\alpha=E_0/(N\Omega)$ and $|\Psi|$ from exact numerics and plotted them as functions of $\Omega/U$. We can see that as $\Omega/U$ increases, $\alpha$ continuously decreases from a large value  to  $\sim 1$, suggesting the change of boson occupation from many higher Landau levels to the lowest one. During this process the vortex core function gradually develops the Gaussian form as in (\ref{Gaussian}), i.e., $\ln(|\Psi|/\rho)$ linearly depends on $\rho^2/l_{\Omega}^2$ with slope $-1/2$, see Fig.~\ref{fig_LLL}(c).

Before closing, we comment on the validity of local density approximation (LDA) used in the GP equation. 
In the needling regime $\Omega\ll U$, LDA is well justified since  the interaction energy ($\sim U$) is much larger than the noninteraction part ($E_0/N=\alpha \Omega$). This can be seen from Fig.~\ref{fig_LLL}(b), for instance, at small $\Omega/U(\ll0.5)$ we have  $\alpha\lesssim 2 \ll U/\Omega$. As $\Omega/U\to0$, the density inhomogeneity should play the same role as it does in an individual static droplet with size $\sim l_\Omega$ and atom number $\sim \nu=N/N_{\nu}$. Given very large $\nu(>10^4)$ in the needling regime (see the Appendix), the density inhomogeneity can be neglected and the LDA can be validated. However, in the opposite limit with $\Omega/U\gtrsim 1$, the kinetic energy contributed from density inhomogeneity may not be always less than $U$. Therefore, the treatment of LDA and the LHY correction calculated from a homogeneous system may not be very accurate in this regime and needs further examination.

\section{Conclusion}
In summary, we have demonstrated a self-bound and clearly visible triangular vortex lattice in a rapidly rotating 2D Bose-Bose droplet in the quantum Hall limit. The revealed distinct structures of vortex lattice (see Fig.~\ref{fig_schematic}) and the smooth crossover in between, as well as the flat-top surface and its universal dependence on  $\Omega/U$ [Fig.~\ref{fig_needling}(b)] can be readily detected in the current experiments of ultracold droplets. 
Intriguingly, here the quantum fluctuations (or LHY corrections), rather than melting the vortex lattice as in the gaseous state of rotating bosons \cite{review2,Sinova2002:QuantumMeltingAbsence}, help to stabilize it in a rapidly rotating droplet. In this way,  
our results suggest  the quantum droplet as a fascinating platform for the realistic exploration of the quantum Hall regime with rotating bosons. In the future, it will be interesting to investigate how the strongly correlated quantum Hall states emerge in this platform when the filling factor $\nu$ gets sufficiently low, as previously studied for single-species bosons \cite{Cooper, Sinova2002:QuantumMeltingAbsence,Regnault2003:QuantumHallFractions}.

\acknowledgments
We thank Hui Zhai for valuable discussions.
The work is supported by the National Key Research and Development Program of China (Grant No. 2018YFA0307600), the National Natural Science Foundation of China (Grants No. 12074419 and No. 12134015), and the Strategic Priority Research Program of Chinese Academy of Sciences (Grant No. XDB33000000).

\appendix

\section{Imaginary time evolution of  extended GP equation.}

We take a large square box with size $30\times30 (l_{\Omega}^{2})$, which  is further divided into a mesh grid of $512\times512$ for the discretization of wavefunction. The imaginary time step generally takes the value $\Delta \tau=10^{-4}\sim10^{-5} (1/\omega)$. 

We have used different methods for the iterations in imaginary time, such as split-step finite difference  \cite{Antoine2013:ComputationalMethodsDynamics} and the split-step Fourier method  \cite{Antoine2013:ComputationalMethodsDynamics}. Throughout the iterations, the atom number remains fixed. We have also tried various initial states, such as the single vortex state and the Gaussian function with a random phase $\Phi(x,y)$. 
 We have confirmed that the final steady state is always the self-bound triangular vortex lattice. For accuracy and efficiency of the iteration, we finally adopt the split-step Fourier method to obtain  large vortex lattices (with $120\sim130$ vortices) as presented in this work. 

\begin{figure}[ht]
    \centering
    \includegraphics[width=8.2cm]{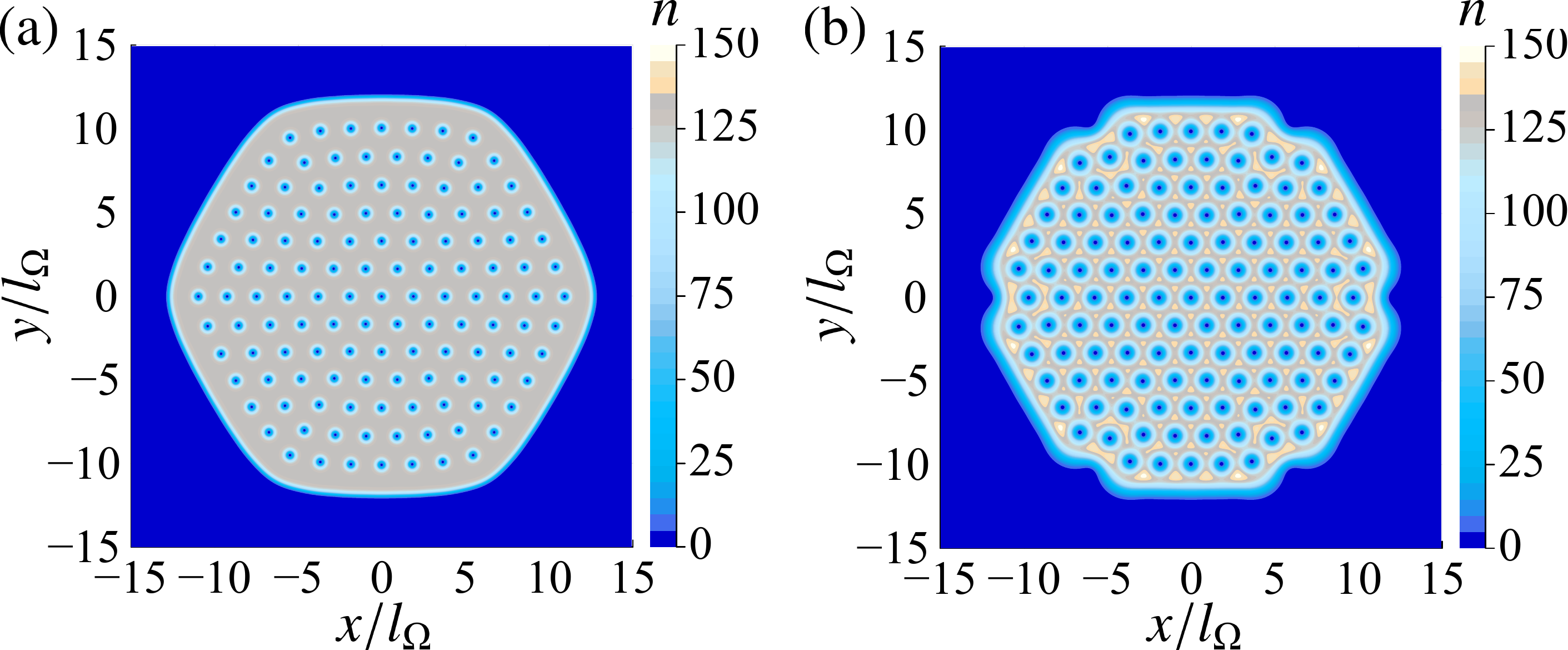}
    \caption{Real space density distributions of vortex lattices in Figs.~\ref{fig_needling}(a) and \ref{fig_LLL}(a) in the main text. The color scale labels the density in units of $(\mu \text{m})^{-2}$. }\label{fig5}
\end{figure}

For the initial state $\Psi_{0}$, we take the two following steps in order to achieve the fast convergence.  First, we turn off the trapping potential and rotating terms, and input a Gaussian state to evolve. This step generates a static droplet $\Psi_\text{droplet}$ in vacuum. Second, we perform phase ``imprinting'' on the droplet  \cite{Ancilotto2017:DensityFunctionalTheory} to prepare $\Psi_0$,
$$
\Psi_{0} = \Psi_{\text{droplet}}\cdot \prod_{j} \dfrac{z-z_j}{|z-z_j|},
$$
where the complex coordinate is $z=x+iy$, and $\{z_{j}\}$ are vortices' initial positions. Here we make use of $C_{6}$ symmetry and the Feynman-Onsager relation to select $\{z_{j}\}$ as triangular lattice sites,
$$
z_{j}/l_{0}=(2m_{1}+2m_{2}\times e^{i\pi/3})\times e^{i m_{3} \pi/3},
$$
where $l_{0}=\sqrt{\pi/(2\sqrt{3})}l_{\Omega}$ is the theoretical value of half spacing between the nearest vortices, $m_{1}$ and $m_{2}$ are integers, and $m_{3}\in\{0,1,2,3,4,5\}$. For example, $m_{2}$ take values from 0 to 6, and $m_{1}$ from 0 to $6-m_{2}$. Using this initial state and time step $10^{-5}$, we obtain the converged state
after $\sim 10^{6}$ iterations. Typical distributions of vortex lattices are shown in Fig.\ref{fig5}.

\begin{figure}[ht]
    \centering
    \includegraphics[width=7.cm]{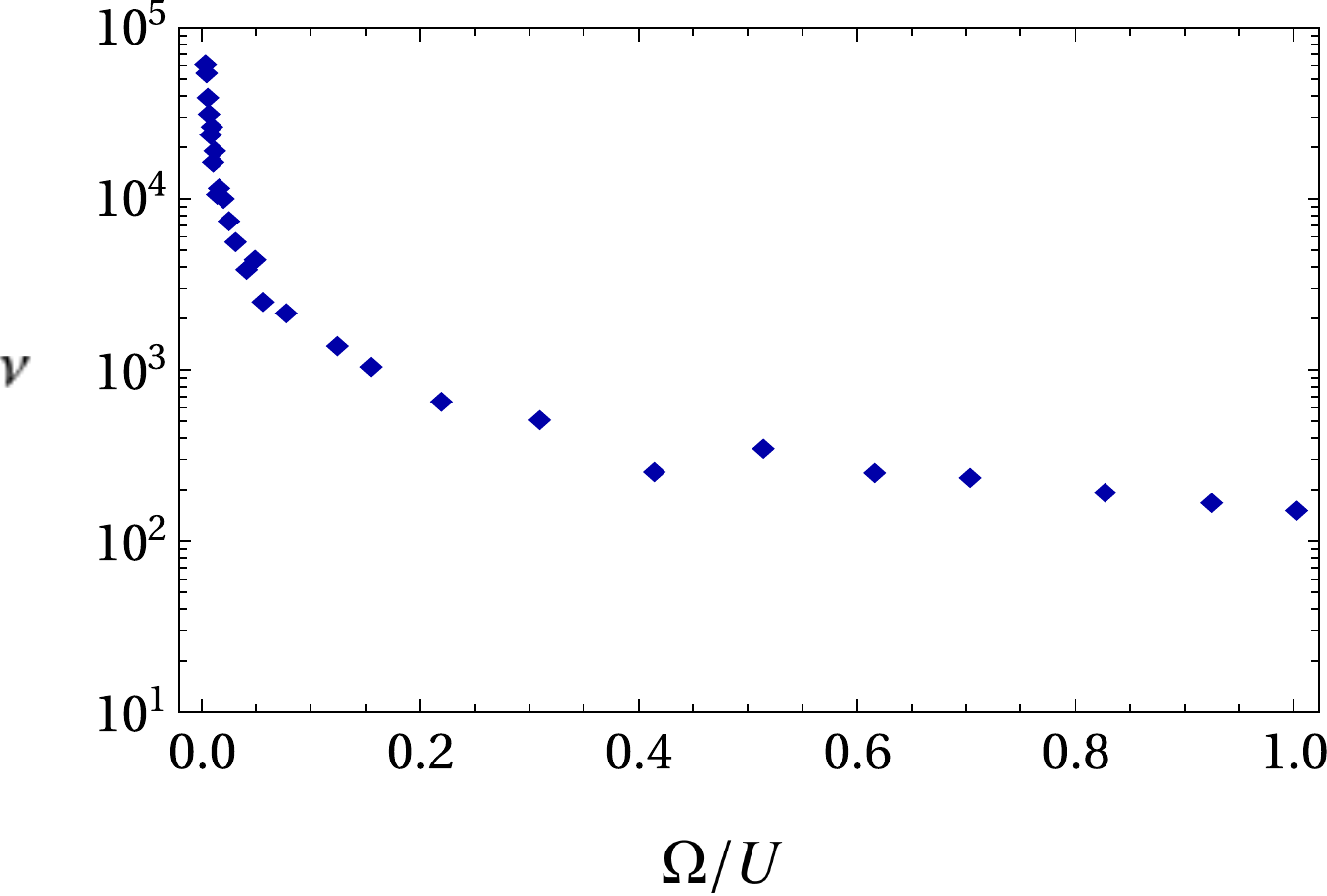}
    \caption{Filling factor $\nu=N/N_v$ as a function of $\Omega/U$}.\label{fig6}
\end{figure}

The vortex lattices presented in this work exhibit high filling factor $\nu$ ranging from $\sim 63000$ to $\sim150$, see Fig.~\ref{fig6}. Various physical parameters of two typical samples (at two different $\Omega/U$) are shown in Table \ref{table}.

\begin{table}[H]
    \centering
    \begin{tabular}{ c |c |c |c| c| c| c| c }
        $\Omega/U$ & N & $\nu$ & $\alpha$ & $\mu$ & E/N & $\sqrt{\braket{r^2}}$ & $\bar{n}/n_0$ \\ 
        \hline
        0.03 & $734000$ & $5780$ & 1.30 & -30.7345 & -30.7334 & 8.68 & 1.05 \\
        1.00 &  $18800$ & $155$ & 1.00 & 0.04546 & 0.04518 & 8.43 & 1.21  
    \end{tabular}
    \caption{Parameters of vortex lattices at two specific values of $\Omega/U$. Here $N$ is the particle number, $\nu$ is the filling factor, $\alpha$ is the non-interacting energy per particle, $\mu$ is the chemical potential, and $\sqrt{\braket{r^2}}$ is the root-mean-square radius. Here the length and energy units are, respectively, $l_{\Omega}$ and $\Omega$. } \label{table}
\end{table}

\end{document}